\documentclass[aps,prb,twocolumn,showpacs,superscriptaddress,reprint]{revtex4-1}
\usepackage{graphics}
\usepackage{xcolor}   
\usepackage{amsmath}
\usepackage{epsfig}
\usepackage{verbatim}

\newcommand{\ih}{\frac{i}{\hbar}}

\newcommand{\ibl}{\mathcal{B}^L}

\renewcommand\vec[1]{\overrightarrow{#1}}
\newcommand{\ep}{\varepsilon}
\newcommand{\cVV}{\Phi}
\newcommand{\cIV}{\vec{\Psi}}

\newcommand{\up}{{R_1}}
\newcommand{\low}{{R_2}}

\newcommand{\Unit}{\Delta}
\begin{document} 
\title{Three-terminal interface as a thermoelectric generator beyond Seebeck effect}
\author{Hangbo Zhou}
\affiliation{Institute of High Performance Computing, A*STAR, 138632, Singapore}
\author{Gang Zhang}
\email[]{zhangg@ihpc.a-star.edu.sg}
\affiliation{Institute of High Performance Computing, A*STAR, 138632, Singapore}
\author{Jian-Sheng Wang}
\affiliation{Department of Physics, National University of Singapore, 117551, Singapore}
\author{Yong-Wei Zhang}
\email[]{zhangyw@ihpc.a-star.edu.sg}
\affiliation{Institute of High Performance Computing, A*STAR, 138632, Singapore}

\date{\today}

\begin{abstract}
We investigate thermoelectric transport through interfaces with inelastic scatterings by developing a quantum theory, which has been extensively validated by existing theories. We find that under temperature bias, while a two-terminal conductor-insulator interface behaves only as a thermal resistor, a three-terminal conductor-insulator-conductor interface can function as an electricity generator caused by phonon-mediated electron scatterings with heat-charge current separation. Unlike conventional thermoelectrics which is a property of a bulk caused by the Seebeck effect, this thermoelectric behavior is a property of an interface driven by electron-phonon scatterings.

\end{abstract}
\maketitle



Thermoelectrics is a physical process that converts heat energy to electricity. In the last few decades, huge efforts have been made in searching for high-performance thermoelectric materials by maximizing a dimensionless quantity called figure of merit \cite{Dubi2011}. The figure of merit is a near-equilibrium calibration of thermoelectric performance when the underlying mechanism is the Seebeck effect \cite{goldsmid2016,Rowe1995}, which describes that a voltage bias between two ends of a material is induced by a temperature bias along the same channel due to asymmetric diffusion of electrons above and below Fermi-energy. The material itself acts as an energy filter that blocks electrons through energy gap or unevenly distributes carrier density and mobility \cite{Benenti2017}. However, the challenge in this approach is that the factors that determine the figure of merit, such as the Seebeck coefficient, electric conductivity and thermal conductivity, which are measured in the same channel, are often interrelated. For example, semiconductors with a suitable bandgap are able to produce high Seebeck coefficients but low electric conductivity. Another example is that the thermal conductivity and electric conductivity are often proportional to each other (The Wiedemann-Franz law).

Along with the advances in nanotechnology, explorations of other thermoelectric mechanisms beyond two-terminal Seebeck effect become appealing. When the device length scale is smaller than the electronic relaxation length, the electrons become far from local equilibrium and the behavior of electrons is no longer just diffusive \cite{Hung2016}. Instead, the detailed scattering mechanisms become important, which provides a rich playground for new physics of thermoelectrics. For instance, in a triple quantum dots system, it has been demonstrated that heat and charge currents can be separated in a thermoelectric device and the performance is enhanced through separate controls of electric and thermal conductivities\cite{Mazza2015}. Adding a phonon bath via electron-phonon interaction to double quantum dots system as heat supply is also shown to be favourable for thermoelectrics \cite{Entin2010,Jiang2012}. Adding a heat exchange terminal through electrons to a thermoelectric system has also been proposed such as Aharonov-Bohm rings\cite{Balachandran2013} and voltage probes with broken time-reversal symmetry\cite{Brandner2013,Brandner2013b}.

 In this work, we propose that a weekly coupled three-terminal interface is able to function as a thermoelectric module that goes beyond the Seebeck effects, due to the extensive momentum and energy exchanges at the interface.
 We demonstrate that while a two-terminal conductor-insulator interface only behaves as a heat resistor, a three-terminal conductor-insulator-conductor interface can behave as a thermoelectric generator. The electron-phonon scatterings at the interface are found to be an active component that drives the thermoelectrics.
 Furthermore the separation of heat and charge currents is obtained as the heat flows from the insulator to both conductors but the electric current flows between the conductors only. The interface scatterings play the role as an energy filter via the selection rule of electron-phonon interactions. Therefore the achieved thermoelectrics becomes a property of interface itself. As a result, in addition to material search, it suggests a new direction of thermoelectrics through interface engineering. 

In order to investigate transport through the interfaces, a suitable quantum transport theory is necessary. However, the established theories face challenges in handling interfaces. For instance, the non-equilibrium Green's function (NEGF) is difficult to handle inelastic scatterings, which are exceedingly important at interfaces. The breakdown of periodicity at interface also restricts the use of Boltzmann transport equation in momentum space.  Therefore, we have developed a theoretical framework from a Hamiltonian level, which is specific to solve interface transport problems. It is general enough to handle most types of transport carriers including both electrons and phonons. In this theory, the interface is modelled by connecting a left Hamiltonian $H^L$ to a right one $H^R$ via interfacial coupling $H^{LR}$ for generality. The total Hamiltonian can be written as $H=H^L+H^R+H^{LR}$. The interface Hamiltonian follows the form of direct product of coupling operators 
\begin{equation}
H^{LR}=\sum_{\alpha\beta}V_{\alpha\beta}B^L_\alpha\otimes B^R_\beta,
\end{equation}
where $B^L(B^R)$ is a column vector of the operators belonging to the left (right) region that couples to the right (left) region and $V$ is a matrix of the coupling strength. 

Currents are the observables of interest in studying thermoelectrics. The current flow from one region to the other is related to conserving quantities within each region. Hence the current operator can be defined as the rate of change for that conserving quantity in one region. For instance, the energy current operator can be defined as the derivative of left region Hamiltonian, and electric current operator can be defined from the change rate of electron number in one region. To be general, we use $X^L$ to represent such conserving quantity in the left region, and then the current operator flowing out of the left region can be defined as its time derivative
\begin{equation}
I^L=-\frac{dX^L}{d\tau}=-\ih\sum_{\alpha\beta}V_{\alpha\beta}[B_\alpha^L,X^L]\otimes B_\beta^R.
\end{equation}
By replacing $X^L$ with $H^L$, one obtains the energy current. Similarly, by replacing $X^L$ with electron number operator $N^L$, one obtains the particle current. 

In the regime where the coupling matrix $V$ is small, one can derive that in steady state, the expectation value of current operator can be written as
\begin{equation}
\label{eq:major}
\bar{I}^L=-\ih\int_{-\infty}^{\infty}d\tau\,\mbox{Tr}\big\{\cIV^L(\tau)V[V\cVV^R(\tau)]^T\big\},
\end{equation}
where both $\cIV^L(\tau)$ and $\cVV^R(\tau)$ are the matrices of correlation functions. Explicitly they are
\begin{equation} 
\cIV^L(\tau)=\mbox{Tr}[\rho^L\ibl(\tau)(B^L)^T],
\end{equation}
\begin{equation}
\cVV^R(\tau)=\mbox{Tr}[\rho^RB^R(\tau)(B^R)^T],
\end{equation}
where $\ibl_\alpha$ is a shorthand notation of $-\ih[B_\alpha^L,X^L]$, $\rho^L(\rho^R)$ is the density matrix of left(right) region and the time evolution is according to $H^L$ or $H^R$, in other words, $\ibl(\tau)=e^{iH^L\tau/\hbar}\ibl e^{-iH^L\tau/\hbar}$ and $B^R(\tau)=e^{iH^R\tau/\hbar}B^R e^{-iH^R\tau/\hbar}$.
A detailed derivation of Eq.~(\ref{eq:major}) is given in Section A of Supplemental Material. It can be shown that this formalism is accurate to the second order of the coupling strength $V$. In Section B of Supplemental Material, we show two examples of applying this formalism to heat and electronic transport problems for interfaces with only elastic scattering. In order to verify the consistency of this theory from existing theories, we show (1) in Section C of Supplemental Material, that for heat transport of a Rubin chain, this formalism analytically matches with the non-equilibrium Green's function formalism \cite{Li2012} up to the second order of $V$; (2) in Section D of Supplemental Material, for a heat bath coupled to a finite system, this formalism analytically matches with the current formula derived from the quantum master equation approach (QME) \cite{Thingna2012,Zhou2015}. 

This formalism can be further generalized to multi-terminal interfaces where the total Hamiltonian becomes $H=H^1+H^2+\cdots+H^n+H^{int}$ and interface coupling Hamiltonian becomes $H^{int}=\sum_{\{\alpha_i\}}V_{\alpha_1\cdots\alpha_n}B^1_{\alpha_1}\otimes B^2_{\alpha_2}\otimes \cdots B^n_{\alpha_n}$, where $B^i_{\alpha_i}$ is the $\alpha$th coupling operator belonging to Hamiltonian $H^i$. Through a straightforward derivation, one can find that Eq.~(\ref{eq:major}) can be generalized as 
 \begin{equation}
 \label{eq:multi}
 \bar{I}^i=-\ih\int_{-\infty}^{\infty}\!\!\!\!d\tau\sum_{\{\alpha,\alpha'\}}V_{\alpha_1\cdots\alpha_n}V_{\alpha'_1\cdots\alpha'_n}\cIV^i_{\alpha_i\alpha'_i}(\tau)\prod_{j\ne i}\cVV^j_{\alpha_j\alpha'_j}(\tau)
 \end{equation}
 where $\bar{I}^i$ is the expectation value of the current flowing out of lead $i$ and the summation is over all the combinations of indices $\alpha$ and $\alpha'$.

\begin{figure}
 	\includegraphics[width=0.9\linewidth]{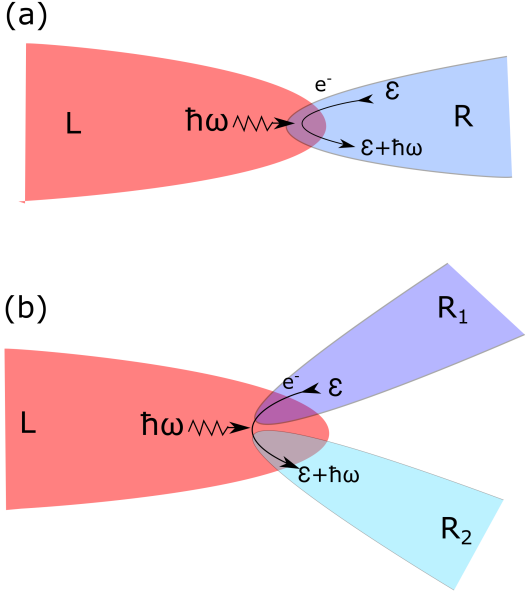}
 	\caption{\label{fig:setup} (a)Thermal transport in two-terminal conductor-insulator interface. The energy of phonons are absorbed by the electrons in the conductor, resulting in a heat flow from left to right region. (b)Thermoelectric transport in conductor-insulator-conductor interface. Two conductors with equal temperature are connected to an insulator. Electrons in one conductor can hop to the other via absorption or emission of phonons from the insulator, resulting thermoelectric current flowing along the two conductors.}
 \end{figure}

Next we apply our theory to an insulator-conductor interfaces where electron-phonon interaction is predominantly through scatterings. We explore two kinds of interfaces. The first one is a two-terminal insulator-conductor interface as shown in Fig.~\ref{fig:setup}(a), where both the pre- and post- scattering electrons are in the same conductor. The second one is a three-terminal setup as shown in Fig.~\ref{fig:setup}(b), where electrons from one conductor can hop towards the other conductor via absorbing or emitting phonons. 

\textit{Two-terminal interfaces} The two-terminal setup can be modelled by a phonon Hamiltonian that is coupled to an electron Hamiltonian via electron-phonon interaction (EPI) as follows
\begin{equation}
\label{eq:ham}
H=H^L+H^R+H^{epi},
\end{equation} 
where $H^L=\sum_{q}\frac{(p_q)^2}{2m_q}+\frac{1}{2}m_q\omega^2_q(x_q)^2$ is the phonon Hamiltonian with $p_q$, $m_q$, $\omega_q$ and $x_q$ being the momentum, mass, frequency and displacement of phonon mode $q$. $H^R=\sum_{k}\ep_k c^\dagger_kc_k$ is the electron Hamiltonian with $\ep_k$, $c^\dagger_k$ and $c_k$ being the energy, creation and annihilation operator of electron with wave-vector $k$. Here we have ignored the phonons in the conductor. The EPI Hamiltonian is written as $H^{epi}=\sum_{nij}M^n_{ij}x_nc^\dagger_ic_j$,
 where $M$ is the coupling tensor containing information of EPI strength. Written in the normal coordinates of $H^L$ and $H^R$, this Hamiltonian becomes $H^{epi}=\sum_{nij,qkl}M^n_{ij}c^n_qv^i_\beta v^j_\gamma x_qc^\dagger_\beta c_\gamma$, where $x_n=\sum_q c_q^nx_q$ and $c_i=\sum_\beta v_\beta^i c_\beta$ are the normalization coefficients. We can verify that the electron number operator of right region, $N_R=\sum_kc^\dagger_kc_k$, commutes with $H$, indicating that the number of electrons in the right region is conserved and the net electronic current flowing through the interface equals zero, independent of the states of the interface. However, the heat current can flow through the interface due to the energy exchange between the electrons and phonons. Indeed, the heat current operator going out of the left region is
\begin{equation}
I_h^L=-\frac{dH_L}{d\tau}=\sum_{nij}M^n_{ij}\frac{p_n}{m_n}c^\dagger_ic_j.
\end{equation}
By plugging these coupling operators into Eq.~(\ref{eq:major}), we found that the heat current flowing out of left region is
\begin{eqnarray}
\label{eq:epi}
\bar{I}_h^L&=&\frac{2}{\hbar^2}\sum_{nij,mkl}M^n_{ij}M^m_{kl}\int\!\!\!\!\int\frac{d\ep_1d\ep_2}{(2\pi)^2}J^L_{nm}(\Omega)n^L(\Omega)\Omega\nonumber\\
&\times& W_{jk\to li}(\ep_1\to\ep_2),
\end{eqnarray}
 where 
 \begin{equation}
 \label{eq:Wij}
 W_{jk\to li}(\ep_1\to\ep_2)=\Gamma^R_{jk}(\ep_1)\Gamma^R_{li}(\ep_2)f^R(\ep_1)[1-f^R(\ep_2)],
 \end{equation}
 is the transmission rate of electrons from energy $\ep_1$ to $\ep_2$. Here $\Omega=(\ep_2-\ep_1)/\hbar$, $J_{nm}(\Omega)=\pi\sum_q \frac{\hbar c_q^nc_q^m}{2m_q\omega_q}\delta(\Omega-\omega_q)$ when $\Omega\ge 0$ and $J(\Omega)=-J(-\Omega)$ when $\Omega<0$; and $\Gamma_{ij}(\ep)=\sum_\alpha 2\pi v^i_\alpha v^j_\alpha \delta(\ep-\ep_\alpha)$ are the phonon and electron spectral density, $n^L(\Omega)=1/(e^{\frac{\hbar\Omega}{k_BT_L}}-1)$ is the Bose-Einstein distribution of phonons and $f^R(\ep)=1/(e^{\frac{\ep-\mu_R}{k_BT_R}}+1)$ is the Fermi-Dirac distribution of electrons containing the information of temperature $T_L$, $T_R$ and chemical potential $\mu_R$.  Equation~(\ref{eq:epi}) is a general formula to calculate the heat current across a conductor-insulator interface mediated by EPI. To verify that this theory obeys the thermodynamic laws, (1) in Section E of Supplemental Material, we show that energy conservation in such interface, explicitly i.e. $\bar{I}^L+\bar{I}^R=0$, is satisfied, which justifies the first law of thermodynamics; and (2) in Section F of Supplemental Material, we show analytically that though the two regions beside the interface have different types of heat carriers, the second law of thermodynamics is strictly guaranteed.

For illustration, we apply this formula to a one-dimensional model that contains a Rubin bath connecting to a tight-binding electron Hamiltonian. The spring constant in Rubin bath is $k$ except for the one near the interface being $k_0$. The electrons have on-site energy $\ep_0$ and hopping energy $t$, which are the diagonal and off-diagonal elements of tight-binding Hamiltonian, respectively. Since only the phonons and electrons near the interface interact with each other, the EPI Hamiltonian can be
 $H_{eph}=M^N_{11}x_Nc^\dagger_1c_1$.
  In this case\cite{Li2012}, the phonon spectral density becomes
 $J(\omega)=-2\mbox{Im}[\frac{\hbar}{k-k_0-k/\lambda_{ph}(\omega)}]$, 
 where 
 $\lambda_{ph}(\omega)=\frac{-\omega'\pm\sqrt{\omega'^2-4k^2}}{2k}$ 
 with $\omega'=(\omega+i\eta)^2-2k/m$.
 The electron spectral density is 
 $\Gamma(\ep)=-2\mbox{Im}[\lambda_e(\ep)/t]$,
 where 
 $\lambda_e(\ep)=\frac{\ep-\ep_0\pm\sqrt{(\ep-\ep_0)^2-4t^2}}{2t}$.
 For both phonons and electrons, the plus or minus sign is determined by the condition 
 $|\lambda_{e(ph)}|<1$. 

\begin{figure}
	\includegraphics[width=\linewidth]{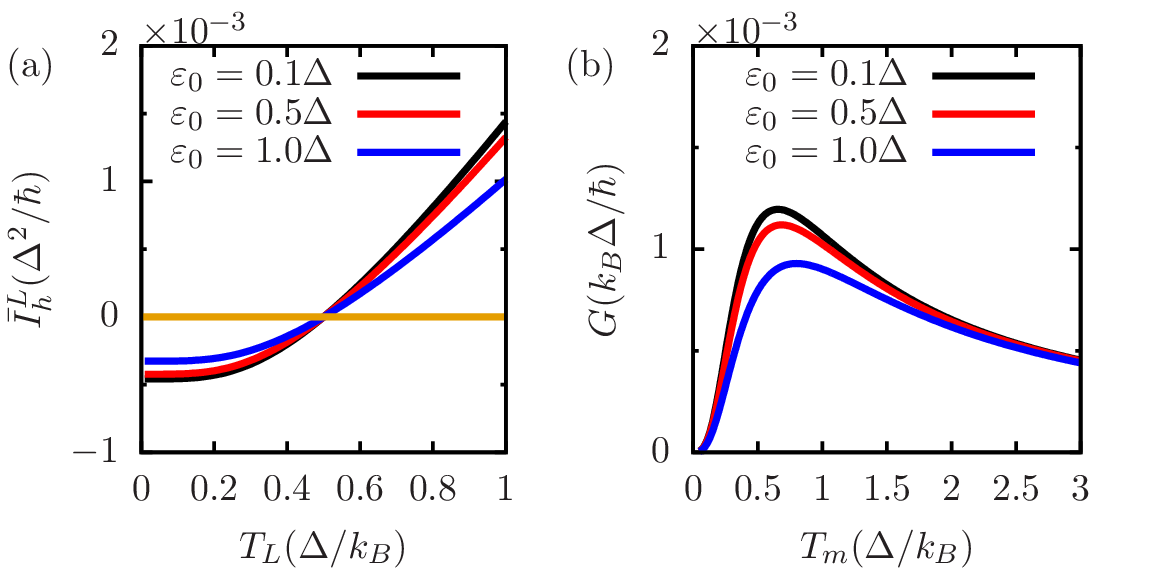}
	\caption{\label{fig:current}(a)Thermal current is plotted against the temperature bias. (b)Thermal conductance is plotted against the average temperature of the leads. We use $\Unit=\hbar\omega=\hbar\sqrt{k/m}$ as reference unit of energy. Then the other parameters are, $t=\Delta$, $\mu=0$, $k_0=k$ and $M^N_{11}=0.1\Delta$. In panel (a) the temperature of right lead is fixed at $T_R=0.5\Delta/k_B$. In panel (b)The conductance is evaluated by $G=\bar{I}^L/(T_L-T_R)$, where $T_{L/R}=T_m\pm 0.01\Delta/k_B$.}
\end{figure}

 By plugging these spectral densities into Eq.~(\ref{eq:epi}), we are then able to calculate the interfacial thermal current. The results shown in Fig.~\ref{fig:current}(a) verify the second law of thermodynamics such that the currents always flow from the high-temperature region to the low one. Interestingly, we find that the temperature dependence of interfacial current exhibits a peak. This peak happens in the regime where the phonon energy spectrum profile fits the electron energy spectrum profile. In the low temperature end, the phonon population is low so that the number of phonons that the electrons can absorb is limited. On the other hand, in the high temperature end, the high-energy phonons are populated and their energy levels go beyond the spectra of electron energy.



\textit{Three-terminal interfaces} In the two-terminal setup, the electric current is zero in any parameter regime. However, in the following, we show that a three-terminal conductor-insulator-conductor interface, as shown in Fig.~\ref{fig:setup}(b), is able to function as a thermoelectric generator. In the setup, the two conductors with equal temperature but different on-site and hopping energies are connected to a phonon lead. Electrons can hop between the conductors via emitting or absorbing phonons. 
This microscopic EPI scattering process is able to cause thermoelectric behavior. For example, when the phonon lead has a higher temperature, the emission of phonons from the insulator can drive electrons hopping in-between the conductors and thus induce an electric current flow between the conductors. Such mechanism is different from the conventional Seebeck effect, since in the conventional Seebeck effect, the influence of temperature is on the Fermi-Dirac distribution of electrons, which causes electron diffusion, rather than interfacial activities.
 
 The Hamiltonian of such problem is similar to Eq.~(\ref{eq:ham}), except that the electron Hamiltonian $H^R$ needs to be split into the upper and lower parts, $H^R=H^\up+H^\low$. Subsequently the EPI Hamiltonian becomes $H^{eph}=\sum_{nij}M^n_{ij}x^L_n[(c^\up_i)^\dagger c^\low_j+(c^\low_j)^\dagger c^\up_i]$. To solve this problem, we need to use the multi-terminal formalism. By applying Eq.~(\ref{eq:multi}), we obtain the expectation value of heat flow out of the phonon lead as
   \begin{eqnarray}
   \label{eq:3Th}
   \bar{I}_h&=&\frac{2}{\hbar}\sum_{nij,n'i'j'}M^n_{ij}M^{n'}_{i'j'}\int\!\!\!\!\int\frac{d\ep_1d\ep_2}{(2\pi)^2}J_{nn'}(\Omega)n(\Omega)\Omega\nonumber\\
   &\times&\Big\{W^{\up\to\low}_{ii'\to jj'}(\ep_1\to\ep_2)+W^{\low\to\up}_{jj'\to ii'}(\ep_1\to\ep_2 )\Big\}.
   \end{eqnarray}
 where $\Omega=(\ep_2-\ep_1)/\hbar$ and the definition of $W$ is the same as Eq.~(\ref{eq:Wij}). We also obtain the electric current flow between the two conductors as
 \begin{eqnarray}
 \label{eq:3Te}
 \bar{I}_e&=&\frac{2e}{\hbar}\sum_{nij,n'i'j'}M^n_{ij}M^{n'}_{i'j'}\int\!\!\!\!\int\frac{d\ep_1d\ep_2}{(2\pi)^2}J_{nn'}(\Omega) n(\Omega)\nonumber\\
 &\times&\Big\{W^{\up\to\low}_{ii'\to jj'}(\ep_1\to\ep_2)-W^{\low\to\up}_{jj'\to ii'}(\ep_1\to\ep_2)\Big\}.
 \end{eqnarray}	
 \begin{figure}
 	\includegraphics[width=\linewidth]{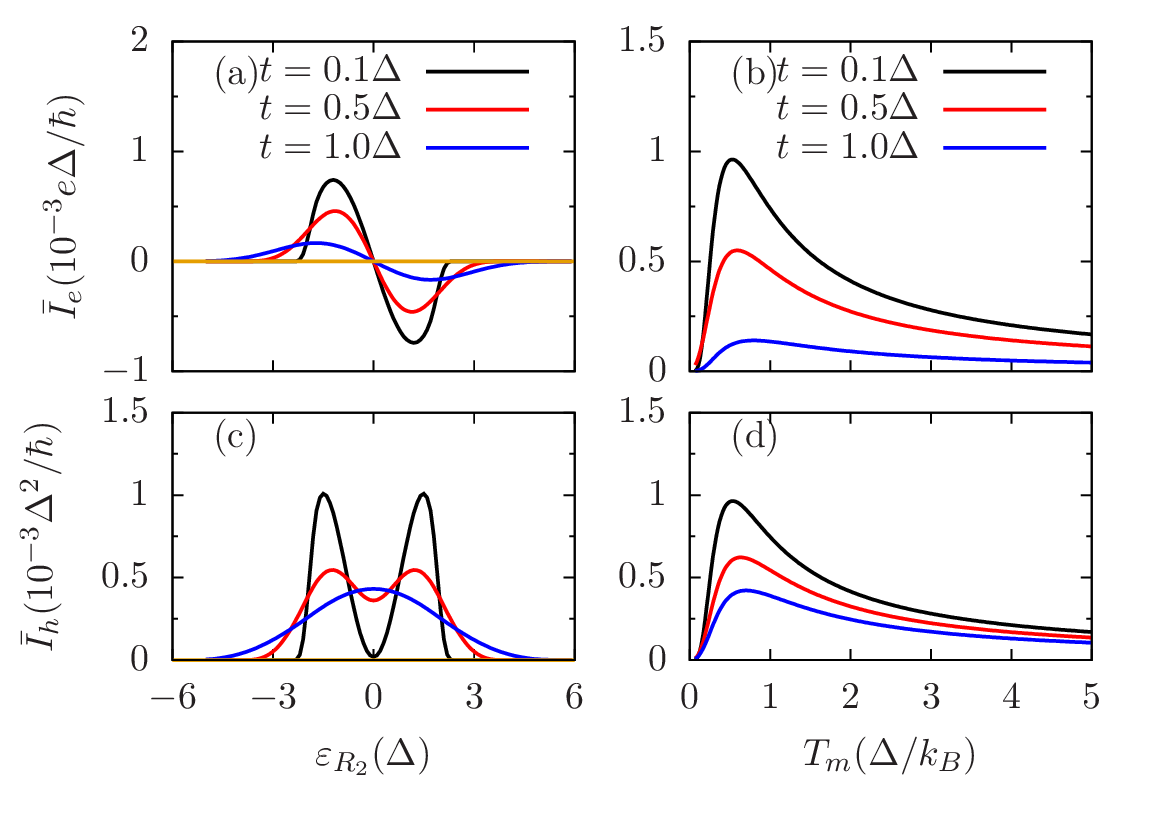}
 	\caption{\label{fig:3Tc}The electric current (a and b) and thermal current (c and d) is plotted against the onsite energy of lead $\up$ (a and c) and the average temperature (b and d). For (a) and (c) we set $T_L=1.1\Delta/k_B$, $T_\up=T_\low=\Delta$ and $\ep_\up=0$. For (b) and (d) we set $T_L=T_m+0.05\Delta/k_B$, $T_\up=T_\low=T_m-0.05\Delta/k_B$, $\ep_\up=1.0\Delta$ and $\ep_\low=0$. All other parameters are the same as those used in Fig.~\ref{fig:current}.}
 \end{figure}

 We then apply this theory to a one-dimensional model. By applying a temperature bias between the phonon lead $L$ and electric leads $\up$ and $\low$, we observe electric current from $\up$ to $\low$.  The models of the leads are the same as before. We break the symmetry between $\up$ and $\low$ by assigning different on-site energies of electrons, namely, $\ep_\up$ and $\ep_\low$, respectively. The results are shown in Fig.~\ref{fig:3Tc}. In panel (a) and (c), we show how thermoelectric currents respond to the asymmetry between lead $\up$ and $\low$. By fixing $\ep_\up$ at 0, the electric current vanishes when the two leads are symmetric ($\ep_\low=0$) and it reaches two peaks at opposite directions when $\ep_\low$ is shifting away from 0. These peaks become smaller with increasing hopping energy. The underlying reason of these peaks is that they occur when the difference of electron energy spectra between lead $\up$ and $\low$ matches the spectrum of phonon lead. Mathematically, it occurs when $\Gamma^\up(\ep_1)\Gamma^\low(\ep_2)$ matches $J(\Omega)$. Under these conditions, the process of electron scatterings between $\ep_1$ and $\ep_2$ through absorbing or emitting a phonon becomes strong. An increase of hopping energy will broaden the electronic energy spectra and hence weaken such electron-phonon scattering process. Figure~\ref{fig:3Tc}(c) shows the heat current flowing from the phonon lead to the two electron leads. Due to the same reason, the heat current also shows two peaks for small hopping energy $t=0.25\Delta$. However, in the symmetry case where $\ep_\low=0$, the heat current does not vanish. This is because the electric current is canceled through the forward and backward scattering processes between lead $\up$ and $\low$, but they both contribute to the heat current. This heat current increases with increasing hopping energy, due to the stronger heat dissipation. Figure ~\ref{fig:3Tc}(b) shows the temperature dependence of interfacial electric current. In the low temperature regime, the interfacial electric current increases with temperature due to the increased number of excited phonons. In the other regime, the high temperature will smooth out the asymmetry of electronic spectra and thus reduce the electric current. As a consequence, there exits an optimized temperature for the interfacial electric current, which is around $T_m=0.7\Delta/k_B$. Similar effects can also be observed for heat current as shown in Fig.~\ref{fig:3Tc}(d).

 \begin{figure}
 	\includegraphics[width=\linewidth]{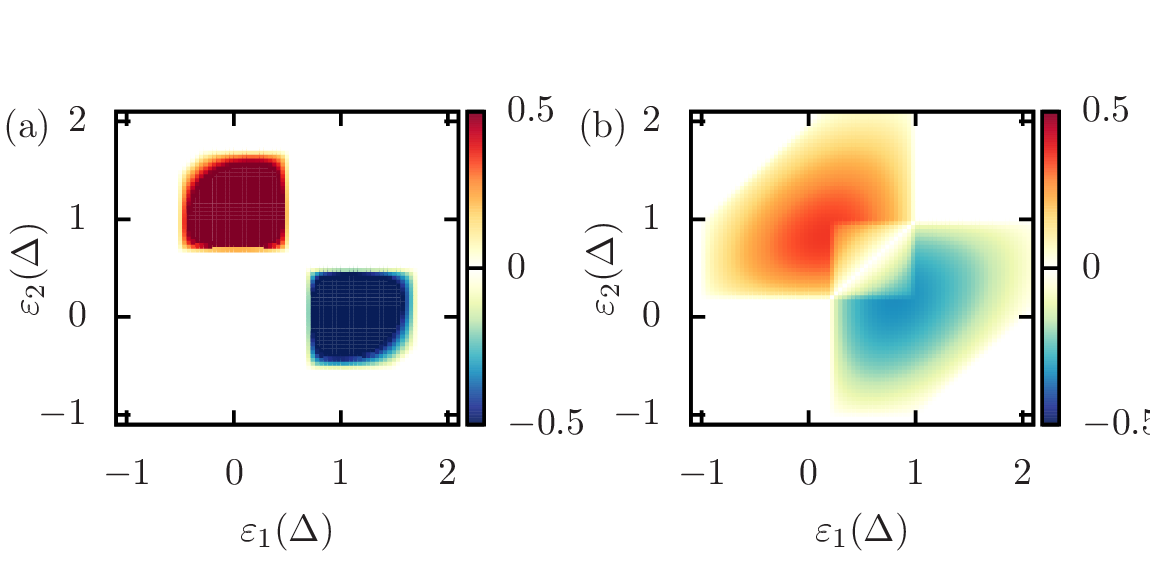}
 	\caption{\label{fig:scatter2}The integrand of Eq.~(\ref{eq:3Te})(including the pre-factor of $2(M^N_{11})^2/\hbar^2$, in units of $10^{-2}/\hbar$) as a function of electron energy $\ep_1$ and $\ep_2$ in case of $t=0.25\Delta$ (a) and $t=0.5\Delta$ (b). For both panels, we set $T_L=1.1\Delta/k_B$, $T_\up=T_\low=1.0\Delta$, $\ep_\up=0$ and $\ep_\low=1.2$. All the other parameters are the same as those used in Fig.~\ref{fig:current}.}
 \end{figure}
 Figure ~\ref{fig:scatter2} shows the scattering probability of electrons between lead $\up$ and $\low$. Electrons with energy in the red region experience forward transport (from $\up$ to $\low$ ) while in the blue regime, they experience backward transport (from $\low$ to $\up$). The overall electric current comes from the asymmetry between these two regimes, which can be seen after numerical integration. However, we can immediately conclude that for a small hopping energy [Fig.~\ref{fig:scatter2}(a)], their scattering spectra are dense and narrow. With increasing hopping energy, the scattering spectra broaden and overlap. Electrons with energies in the overlap regime can transport both forwards and backwards, cancelling out their contributions to the overall electric current. Therefore, it suggests that sharp and dense energy spectra at the scattering interface have better interfacial electric performance.
  
\begin{figure}
 	\includegraphics[width=0.9\linewidth]{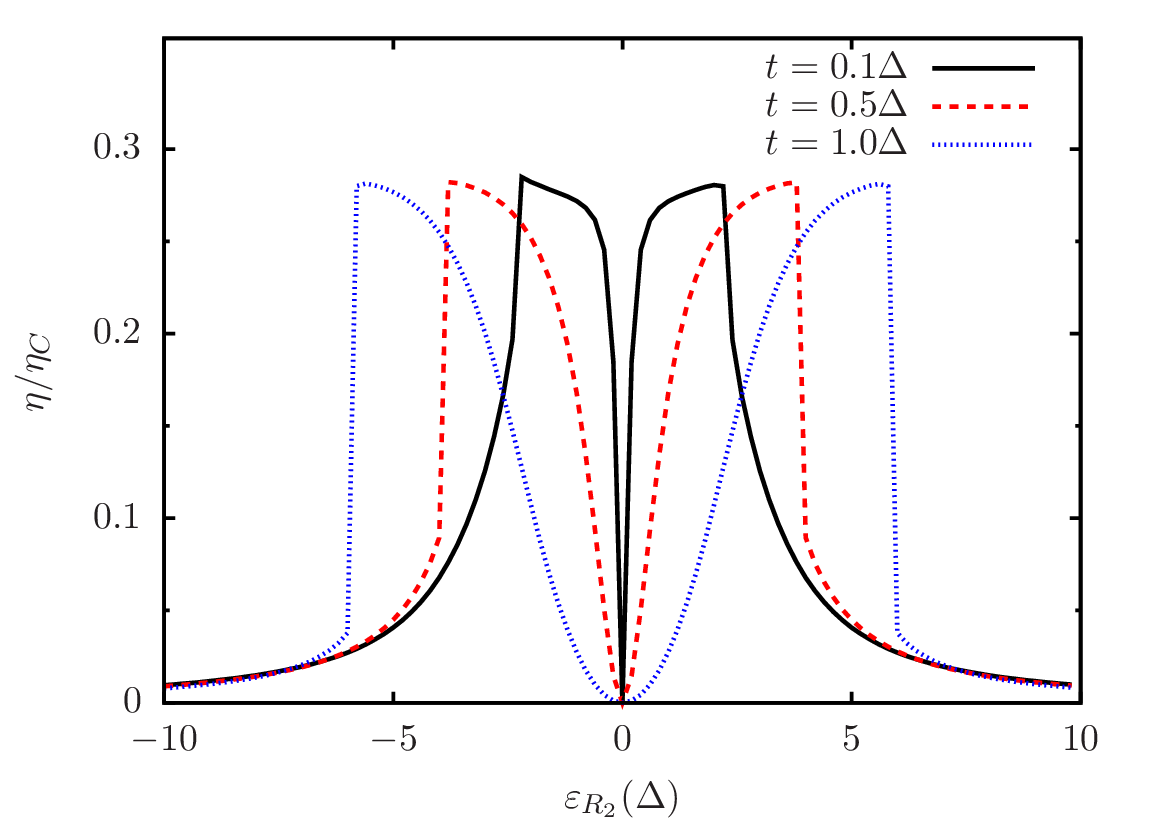}
 	\caption{\label{fig:eta} The efficiency, which is divided by Carnot efficiency, is plotted against $\ep_\low$ under different hopping energies $t$. The parameters are all the same as those used Fig.~\ref{fig:3Tc}(a).}
 \end{figure}

 Next we discuss the heat-work conversion efficiency of the three-terminal thermoelectric generator. We analyze the efficiency in the framework established by Goldsmid\cite{goldsmid2016}. Suppose an external load with resistance $R$ is connected to this thermoelectric generator, the efficiency is measured as the work done on the load divided by the heat supply from the phonon bath. Due to the effect of the external load, the electric current will be reduced by a factor of $R'/(R+R')$, where $R'$ is the internal resistance of the generator. Therefore the work done on the load is $W=\frac{I_e^2R'^2R}{(R+R')^2}$. Due to the heat-charge current separation, we expect that the heat current from the insulator $I_h$ is not affected by the external load. As a result, the efficiency is given by $\eta=I_e^2R'^2R/[I_h(R+R')^2]$. In this case, the maximum efficiency and maximum power are reached simultaneously when $R=R'$, resulting in maximum efficiency $\eta=I_e^2R'/(4I_h)$. The internal resistance $R'$ can be calculated via $R'=(\mu_\up-\mu_\low)/I_e$ by applying a small bias $\mu_\up-\mu_\low=0.001\Delta$ to the two conductors and keeping all the three terminals at the same temperature. The results of the obtained efficiency are shown in Fig.~\ref{fig:eta}. It is seen that for all the hopping energies, the maximum efficiency can reach $0.28\eta_C$, where $\eta_C$ is the Carnot efficiency. For conventional thermoelectric materials such high efficiency can only be reached when the figure of merit is $2.16$, showing a great potential of this three-terminal generator to achieve a high-efficiency. In the symmetric point $\eta_\low=0$, the efficiency is 0 due to the lack of electric current. Efficiency peaks appear at both sides of the symmetry point. With increasing hopping energy $t$, the positions of the peaks shift outwards due to the broadening transmission spectrum. We notice that when $t=0.1\Delta$, the regime of maximum efficiency matches well with the regime of maximum current $I_e$ and the output power, distinctively different from conventional thermoelectric materials that the maximum efficiency and power are always disjoint. 

In conclusion, we have developed a theoretical framework that enables the analysis of interfacial transport properties. This theory has been extensively validated by existing theories, such as NEGF, quantum master equations, and the first and second laws of thermodynamics. We find that, a three-terminal conductor-insulator-conductor interface can behave as a thermoelectric generator. The emission of phonons at phonon lead can drive the motion of electrons between the two conductors with equal temperature. We show that such three-terminal thermoelectric generator has a large heat-work conversion efficiency. Hence, in addition to searching for high performance thermoelectric bulk materials,  our work suggests an alternative route, i.e. investigating thermoelectrics through interface engineering.

\bibliography{interface}

\end{document}